\documentclass[toc]{PoS}
\usepackage{natbib}
\usepackage{amssymb}
\usepackage{amsmath}
\usepackage{xspace} 
\usepackage{graphicx} 

\def\IGR{IGR\,J17544-2619}
\def\xte{XTE\,J1739-302}
\def\igg{IGR\,J16479-4514}
    
\def\igrjj{IGR\,J16465-4507} 
\def\igrj11{IGR\,J11215-5952}

\def\2s{2S\,0114+65}

\def\chan{{\em Chandra}}
\def\xmm{{\em XMM-Newton}}
\def\int{{\em INTEGRAL}}

\title{Hunting for Magnetars in High Mass X-ray Binaries. The Case of SuperGiant Fast X-Ray Transients}

\ShortTitle{Magnetars and SFXTs}

\author{\speaker{E. Bozzo}$^{a,b}$ M. Falanga$^{c}$ L. Stella$^{a}$\\
\llap{$^{a}$}INAF - Osservatorio Astronomico di Roma, Via Frascati 33, 00044 Rome, Italy \\ 
\llap{$^{b}$}Dipartimento di Fisica - Universit\`a di Roma ``Tor Vergata'', via della Ricerca Scientifica 1, 00133 Rome, Italy\\
\llap{$^{c}$}CEA Saclay, DSM/DAPNIA/Service d'Astrophysique (CNRS FRE 2591), F-91191, Gif sur Yvette, France\\}
    
\abstract{In this paper we summarize some aspects of the wind accretion theory in high mass X-ray binaries hosting a magnetic
neutron star and a supergiant companion. In particular, we concentrate on the different types of interaction between the inflowing
wind matter and the neutron star magnetosphere that are relevant when accretion of matter onto the neutron star surface is largely
inhibited; these include inhibition by the centrifugal and magnetic barriers. 
We show that very large luminosity swings ($\sim$10$^4$ or more on time scales as short as hours) can result from transitions across
different regimes. This scenario is then applied to the activity displayed by supergiant fast X-ray transients 
(SFXTs), a new class of high mass X-ray binaries in our galaxy recently discovered with INTEGRAL. 
According to this interpretation we argue that SFXTs which display very large luminosity swings and host a slowly spinning
neutron star are expected to be characterized by magnetar-like fields. 
Supergiant fast X-ray transients might thus provide a unique opportunity to detect and study accreting magnetars in binary systems.}

\FullConference{7th INTEGRAL Workshop\\
		 September 8-11 2008\\
		 Copenhagen, Denmark}

\begin{document}

\section{Introduction}
\label{sec:intro}

Supergiant Fast X-ray transients (SFXTs) are a new class of 
high mass X-ray binaries (HMXB), recently discovered with \int.\ 
These sources are observed to exhibit sporadic outbursts, lasting from minutes to hours, 
with peak X-ray luminosities between $\sim$10$^{36}$ and 10$^{37}$~erg~s$^{-1}$ 
\citep{sguera05}. 
No firm orbital period measurement has been obtained 
yet\footnote{Only IGR\,J11215-5952    
showed recurrent flaring activity, with a periodicity of $\sim$165~d. 
This is interpreted as outbursts from a systems with an unusually long orbital 
period \citep{sidoli07}. Thus \citet{walter07} excluded these sources from their SFXT list.}. 
A recent list of confirmed ($\sim$5) and candidate ($\sim$6)  
SFXTs is given by \citet{walter07}. 
Between outbursts, SFXTs remain in quiescence with 
luminosities in the range $\sim$10$^{31}$-10$^{33}$~erg~s$^{-1}$. 
In some cases, very high peak-to-quiescence X-ray luminosity swings 
(factor of $\sim$10$^{4}$-10$^{5}$) were seen    
on timescales comparable to the outburst duration. 
Some SFXTs showed also flare-like
activity at intermediate luminosity levels \citep[e.g.,][]{zand05}.  
Optical identifications of SFXTs show that these sources are
associated to OB supergiant companion stars \citep[see e.g.,][and 
reference therein]{walter07}. These stars have typically a 
mass of M$_{*}$$\sim$30M$_{\rm \odot}$, optical luminosity of 
log (L$_{*}$/L$_{\odot}$)$\sim$5-6, mass loss 
rate of $\dot{M}_{\rm w}$=10$^{-7}$-10$^{-5}$~M$_{\odot}$~yr$^{-1}$,  
wind velocity of v$_{\rm w}$$\sim$1000-2000~km~s$^{-1}$, and are persistent soft X-ray 
sources with luminosity around $\sim$10$^{32}$~erg~s$^{-1}$ \citep{cassinelli81}.  

It is widely believed that SFXTs contain  
neutron stars (NS) sporadically accreting matter from a supergiant companion, and  
in the prototypical SFXTs 
XTE J1739-302 \citep{sguera06} and IGR J16479-4514 \citep{walter06} 
some evidence has been reported for spin periods in the 
$\sim$1000-2000~s range. 
Proposed models for SFXTs generally involve accretion onto a NS 
immersed in the clumpy wind of its supergiant companion \citep{zand05,walter07}. 
However, if accretion onto the collapsed object of SFXTs takes place both 
in quiescence and outburst, then the corresponding X-ray luminosity 
swing, typically a factor of $\sim$10$^4$-10$^5$, would require  
wind inhomogeneities with a very large density and/or velocity contrast   
\citep[according to the standard wind accretion, the mass capture 
rate onto the NS scales like $\dot{M}_{\rm w}$v$_{\rm w}^{-4}$][]{davidson}. 
Models involving accretion of extremely dense clumps are still being 
actively pursued \citep{negueruela08}.   
The requirement on the density and/or velocity contrasts 
in the wind can be eased if there is a {\it barrier} 
that remains closed during quiescence, halting
most of the accretion flow, and opens up 
in outbursts, leading to direct accretion. 
After reviewing briefly the theory of wind accretion in HMXBs 
(\S~\ref{sec:model} and \ref{sec:trans}), we apply barrier scenarios to 
SFXTs in \S~\ref{sec:results}  and conclude 
that if SFXTs host slowly rotating NSs  
(spin periods of several hundreds to thousands seconds), then they 
must possess magnetar-like fields ($\sim$10$^{14}$-10$^{15}$~G). 

\begin{figure*}[t!]
\centering
\includegraphics[height=4.5 cm]{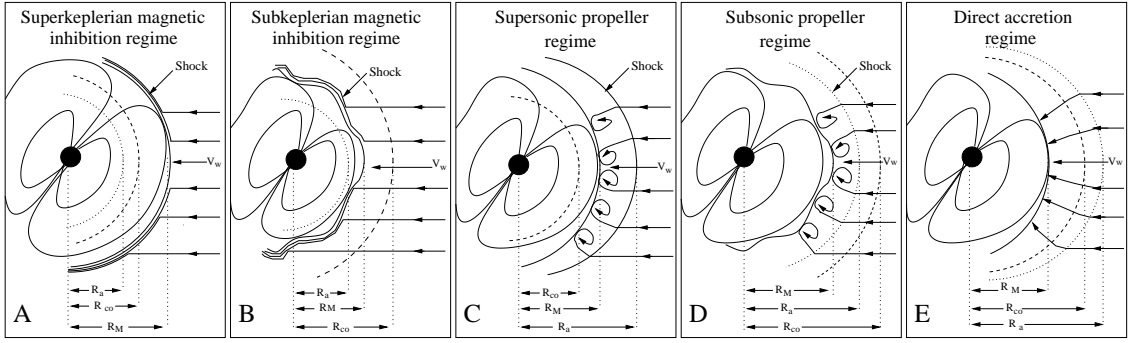}
\smallskip
\caption{\scriptsize Schematic view of a magnetized NS interacting with the 
inflowing matter from its supergiant companion. All the regimes described in the  
text are shown, together with the relative position of the magnetospheric 
radius (solid line), the corotation radius (dashed line), and the accretion 
radius (dotted line). A wavy solid line is used when the magnetospheric 
boundary at the magnetospheric radius is Kelvin-Helmholtz unstable. In the supersonic and subsonic 
propeller regime convective motions at the base of the atmosphere are 
represented with small eddies.}
\label{fig:model} 
\end{figure*}

\section{Stellar wind accretion}
\label{sec:model} 

We investigate here the conditions under which a magnetized 
NS can accrete matter from the 
wind of a massive companion. 
In the theory of wind accretion in HMXBs, the following radii are defined
\citep[e.g.,][]{illarionov75, stella86}:\\ 
{\bf - The accretion radius}, R$_{\rm a}$ is the distance at which the 
inflowing matter is gravitationally focused toward the NS. 
It is usually expressed as   
\begin{equation}
R_{\rm a}=2GM_{\rm NS}/v_{\rm w}^2= 3.7\times10^{10} v_{8}^{-2} ~{\rm cm},
 \label{eq:ra}
\end{equation}
where v$_{8}$ is the wind velocity in units of 1000~km~s$^{-1}$ and we 
assumed that the orbital velocity of the star 
is negligible \citep{fkr}. Throughout the paper we fix the NS radius and 
mass at R$_{\rm NS}$=10$^6$~cm and M$_{\rm NS}$=1.4~M$_\odot$, respectively. 
The fraction $\dot{M}_{\rm capt}$/$\dot{M}_{\rm w}$ of the stellar wind mass loss rate 
($\dot{M}_{\rm w}$) captured by the NS depends on R$_{\rm a}$ through 
\citep{fkr} 
\begin{equation}
\dot{M}_{\rm capt}/\dot{M}_{\rm w}\simeq R_{\rm a}^2/(4 a^2)=2\times10^{-5} v_{8}^{-4} 
a_{\rm 10d}^{-2}. 
\label{eq:dotmcapt}
\end{equation}
Here a=4.2$\times$10$^{12}$a$_{\rm 10d}$~cm is the orbital separation, 
a$_{\rm 10d}$=P$_{\rm 10d}^{2/3}$M$_{30}^{1/3}$, P$_{\rm 10d}$ is the binary orbital 
period in units of 10 days, and M$_{30}$ is the total 
mass of the binary in units of 30~M$_{\odot}$ (we assumed circular orbits).\\ 
{\bf - The magnetospheric radius}, R$_{\rm M}$, at which the pressure of the NS magnetic 
field ($\mu^2$/(8$\pi$$R_{\rm NS}^6$), with $\mu$ the NS magnetic moment) 
balances the ram pressure of the inflowing matter 
($\rho_{\rm w}$$v_{w}^{2}$). 
In the case in which R$_{\rm M}$$>$R$_{\rm a}$, 
the magnetospheric radius is given by \citep{pringle}  
\begin{equation}
R_{\rm M}=3.3\times10^{10} \dot{M}_{-6}^{-1/6} v_{8}^{-1/6} a_{\rm 10d}^{1/3} 
\mu_{33}^{1/3} ~ {\rm cm}.   
\label{eq:rm}
\end{equation} 
Here we assumed a non-magnetized spherically symmetric wind \citep{elsner1977},  
with density\footnote{We approximated a-R$_{\rm M}$$\simeq$a ,  
which is satisfied for a very wide range of parameters.} 
$\rho_{\rm w}$(R$_{\rm M}$)$\sim$$\dot{M}_{\rm w}$/(4$\pi$a$^2$v$_{\rm w}$), 
a NS dipolar field with $\mu_{33}$=$\mu$/10$^{33}$~G~cm$^{3}$, and 
$\dot{M}_{-6}$=$\dot{M}_{\rm w}$/10$^{-6}$ M$_{\odot}$~yr$^{-1}$. 
In the following sections we discuss the range of applicability of Eq.~\ref{eq:rm}, 
and the regimes in which a different prescription for R$_{\rm M}$ should be 
used. \\ 
{\bf - The corotation radius}, R$_{\rm co}$, at which the NS angular velocity 
equals the Keplerian angular velocity, i.e. 
\begin{equation}
R_{\rm co}=1.7\times10^{10} P_{\rm s3}^{2/3} ~{\rm cm}.    
\label{eq:rco}
\end{equation}
Here $P_{\rm s3}$ is the NS spin period in units of 10$^3$~s. 

Changes in the relative position of these radii
result into transitions across different regimes for the NS 
\citep{illarionov75, stella86}. 
In particular, the accretion radius and magnetospheric 
radius depend on the wind parameters (see Eqs.~\ref{eq:ra} and \ref{eq:rm}),
which can vary on a wide range of timescales (from hours to months).  
Therefore, variations in the wind parameters can cause the  
NS to undergo transitions across different regimes on comparably 
short timescales, thus opening the possibility to explain the 
properties of some classes of highly variable X-ray sources through them.
Below we summarise the different regimes of a 
magnetic rotating NS, subject to a varying 
stellar wind.

{\bf - Outside the accretion radius: the magnetic inhibition of accretion   
(R$_{\rm M}$$>$R$_{\rm a}$)}.  
In systems with R$_{\rm M}$$>$R$_{\rm a}$ the mass flow from the companion star interacts 
directly with the NS magnetosphere without significant gravitational focusing, forming  
a bow shock at R$_{\rm M}$ \citep{hard, toropina}. 
At least in the front part of the shock, i.e. in the region around the  
stagnation point, the whole kinetic energy of the inflowing matter is    
converted into thermal energy.  
The power released in this region is of order  
\begin{equation}
L_{\rm shock}\simeq\frac{\pi}{2} R_{\rm M}^2 \rho_{\rm w} v_{\rm w}^3 = 
4.7\times10^{29} R_{\rm M10}^2 
v_8^2 a_{\rm 10d}^{-2} \dot{M}_{-6}~{\rm erg ~s^{-1}}  
\label{eq:lshock}
\end{equation}
(R$_{\rm M10}$=R$_{\rm M}/$10$^{10}$~cm), 
and is mainly radiated in the X-ray band. 
We distinguish two different regimes of magnetic inhibition of accretion:   
\begin{itemize}
\item{{\it The superKeplerian magnetic inhibition regime}}: R$_{\rm M}$$>$R$_{\rm a}$, 
R$_{\rm co}$.  
In this case the magnetospheric radius is larger than 
both the accretion and corotation radii. 
Matter that is shocked and halted close to R$_{\rm M}$ cannot 
proceed further inward, due to the rotational drag of the NS 
magnetosphere which is locally superKeplerian. 
Since magnetospheric rotation is also supersonic, the interaction 
between the NS magnetic field and matter at R$_{\rm M}$ results in 
rotational energy dissipation and thus, NS spin down. 
This process releases energy at a rate 
\begin{equation}
L_{\rm sd} \simeq \pi R_{\rm M}^2 \rho_{\rm w} v_{\rm w} (R_{\rm M}\Omega)^2\simeq  
3.7\times10^{29} R_{\rm M10}^4 \dot{M}_{-6} a_{\rm 10d}^{-2} P_{\rm s3}^{-2}~{\rm erg ~s^{-1}}, 
\label{eq:lmagnsd}
\end{equation}
which adds to the shock luminosity (Eq.~\ref{eq:lshock}). 

\item{\it The subKeplerian magnetic inhibition regime}: R$_{\rm a}$$<$R$_{\rm M}$$<$R$_{\rm co}$. 
In this case the magnetospheric drag is subKeplerian
and matter can penetrate the NS magnetosphere through 
Kelvin-Helmholtz instabilities \citep[KHI,][]{hard} and 
Bohm diffusion \citep{ikhsanov}.  
The mass inflow rate across R$_{\rm M}$ resulting
from the KHI 
depends on the efficiency factor $\eta_{\rm KH}$$\sim$0.1,  
the shear velocity at R$_{\rm M}$ v$_{\rm sh}$, and the densities $\rho_{\rm i}$ and $\rho_{\rm e}$ inside 
and outside the magnetospheric boundary at R$_{\rm M}$, respectively. 
The luminosity released by accretion of this matter onto the NS is given by   
\begin{eqnarray}
L_{\rm KH} = 3.5\times10^{34} \eta_{\rm KH} R_{\rm M10}^2 a_{\rm 10d}^{-2} \dot{M}_{-6} 
(\rho_{\rm i}/\rho_{\rm e})^{1/2} (1+\rho_{\rm i}/\rho_{\rm e})^{-1}~{\rm erg~s^{-1}}, 
\label{eq:lx_kh}
\end{eqnarray} 
or 
\begin{eqnarray}
L_{\rm KH} = 8.8\times10^{34} \eta_{\rm KH} P_{\rm s3}^{-1} R_{\rm M10}^3 
a_{\rm 10d}^{-2} v_8^{-1} \dot{M}_{-6} 
(\rho_{\rm i}/\rho_{\rm e})^{1/2} (1+\rho_{\rm i}/\rho_{\rm e})^{-1} ~{\rm erg~s^{-1}}, 
\label{eq:lx_kh2}
\end{eqnarray}
if the shear velocity is taken to be the gas velocity in the post-shock region close to R$_{\rm M}$ 
or the rotational velocity of the NS magnetosphere, respectively (for simplicity, we consider throughout this paper  
L$_{\rm KH}$ equal to the largest of the above two values).  
The ratio $\rho_{\rm i}$/$\rho_{\rm e}$ can be calculated from the equation of the mass conservation 
across the KHI unstable layer, i.e.   
\begin{equation}
R_{\rm M}^2 \rho_{\rm e} v_{\rm conv}\simeq R_{\rm M} h_{\rm t} \rho_{\rm i} v_{\rm ff}(R_{\rm M}),  
\label{eq:masscons}
\end{equation}  
where h$_{\rm t}$ is the height of the unstable layer \citep{bur}. 
Throughout this paper we assume h$_{\rm t}$=$R_{\rm M}$ \citep{bozzo08}. 
The contribution of Bohm diffusion to the total mass inflow rate through the magnetosphere in 
the subKeplerian magnetic inhibition regime can be calculated according to \citet{ikhsanov}, and we  
found that it is orders of magnitude smaller than that due to the KHI
over the whole range of parameters relevant to this work.  
Similarly, the contribution to the total luminosity resulting from
the shock at the magnetospheric boundary can be 
neglected in this regime. 
\end{itemize} 
{\bf - Inside the accretion radius: R$_{\rm M}$$<$R$_{\rm a}$}. 
Once R$_{\rm M}$ is inside the accretion radius, matter flowing from the 
companion star is shocked adiabatically at R$_{\rm a}$ 
(this produces an almost negligible contribution to the total luminosity) 
and halted at the NS magnetosphere.  
In the region between R$_{\rm a}$ and R$_{\rm M}$  
this matter redistributes itself into an approximately spherical
configuration  
(resembling an ``atmosphere''), whose shape and properties 
are determined by the interaction between matter and 
NS magnetic field at R$_{\rm M}$ \citep{davies,pringle}. 
A hydrostatic equilibrium ensues when radiative losses inside 
R$_{\rm a}$ are negligible\footnote{We checked this is verified for all the 
case of interest for this paper.}; the atmosphere is 
stationary on dynamical timescales, and a polytropic law of the 
form p$\propto$$\rho^{1+1/n}$ can be assumed for 
the pressure and density of the atmosphere. 
The value of the polytropic index $n$ depends on the conditions at 
the inner boundary of the atmosphere, and in particular on the rate 
at which energy is deposited there. Three different regimes can be 
distinguished: 

\begin{itemize}
\item {\it The supersonic propeller regime:} R$_{\rm co}$$<$R$_{\rm M}$$<$R$_{\rm a}$.  
In this case the rotational velocity of the NS magnetosphere at R$_{\rm M}$ 
is supersonic; the interaction with matter in the atmosphere leads  
to dissipation of some of the star's rotational energy and thus spin-down. 
Turbulent motions are generated at R$_{\rm M}$ which convect this 
energy up through the atmosphere, until it is lost 
at its outer boundary. In this case $n$=1/2, and the balance between the 
magnetic and gas pressure gives 
\begin{equation}
R_{\rm M}\simeq2.3\times10^{10} a_{\rm 10d}^{4/9} \dot{M}_{-6}^{-2/9} 
v_8^{4/9} \mu_{33}^{4/9} ~{\rm cm}.
\label{eq:rmsuperapprox} 
\end{equation}
Matter that is shocked at $\sim$R$_{\rm a}$, reaches the magnetospheric boundary at 
R$_{\rm M}$ where the interaction with the 
NS magnetic field draws energy from NS rotation. According to \citet{pringle}, this 
gives the largest contribution to the total luminosity in this regime      
\begin{eqnarray}
L_{\rm sd} & = & 2\pi R_{\rm M}^2 \rho(R_{\rm M}) c_{\rm s}^3
(R_{\rm M}) \simeq \nonumber \\
&& 5.4\times10^{31} \dot{M}_{-6}
a_{\rm 10d}^{-2} v_8^{-1} R_{\rm M10}^{1/2} 
(1+16 R_{\rm a10}/(3 R_{\rm M10}))^{1/2} ~{\rm erg ~s}^{-1}.  
\label{eq:lspsuper}
\end{eqnarray}
In the above equation R$_{\rm a10}$=10$^{-10}$R$_{\rm a}$
and c$_{\rm s}$(R$_{\rm M}$)=v$_{\rm ff}$ 
(R$_{\rm M}$)=(2GM$_{\rm NS}$/R$_{\rm M}$)$^{1/2}$.  

\item{\it The subsonic propeller regime}: 
R$_{\rm M}$$<$R$_{\rm a}$, R$_{\rm co}$, $\dot{M}_{\rm w}$$<$$\dot{M}_{\rm lim}$. 
The break down of the supersonic propeller regime occurs when 
R$_{\rm M}$$<$R$_{\rm co}$, i.e., when the magnetosphere rotation 
is no longer supersonic with respect to the 
surrounding material. The structure of the atmosphere changes and   
the transition to the subsonic propeller regime takes place. 
Since the rotation of the magnetosphere is subsonic, the atmosphere is roughly 
adiabatic ($n$=3/2), and the magnetospheric radius is approximated by 
\citep{pringle}: 
\begin{equation}
R_{\rm M}\simeq2\times10^{10} a_{\rm 10d}^{4/7} 
\dot{M}_{-6}^{-2/7} v_8^{8/7} \mu_{33}^{4/7} ~{\rm cm}. 
\label{eq:rmsubapprox} 
\end{equation}
In the subsonic propeller regime, the centrifugal barrier 
does not operate because R$_{\rm M}$$<$R$_{\rm co}$,  
but the energy input at the base of the 
atmosphere (due to NS rotational 
energy dissipation) is still too high for matter 
to penetrate the magnetosphere at a rate $\dot{M}_{\rm capt}$ 
\citep{pringle}. 
Nevertheless a fraction of the 
matter inflow at R$_{\rm a}$  is expected to accrete onto the 
NS, due to the KHI and Bohm diffusion\footnote{To our knowledge this is the first 
application of the KHI to the subsonic propeller regime.}.  
We found that, in all the cases of interest to this paper, the KHI provides the 
largest contribution to the total luminosity in the subsonic propeller regime. 
By using similar arguments to those above, we estimated in this case   
\begin{equation}
L_{\rm KH} = 
1.8\times10^{35} \eta_{\rm KH} R_{\rm M10}^3
\dot{M}_{-6}  
\frac{(1+16 R_{\rm a10}/(5 R_{\rm M10}))^{3/2} 
(\rho_{\rm i}/\rho_{\rm e})^{1/2}}
{P_{\rm s3} a_{\rm 10d}^{2} v_8 (1+\rho_{\rm i}/\rho_{\rm e})} ~{\rm erg ~s}^{-1}. 
\label{eq:lkhsub} 
\end{equation}  
The subsonic propeller regime applies until the critical accretion rate 
\begin{equation}
\dot{M}_{\rm lim_{-6}} = 2.8\times10^2 P_{\rm s3}^{-3} 
a_{\rm 10d}^{2} v_8 R_{\rm M10}^{5/2} (1+16 R_{\rm a10}/(5 R_{\rm M10}))^{-3/2}
\label{eq:dotmlimsub} 
\end{equation}
is reached, at which the gas radiative cooling (bremsstralhung) completely damps  
convective motions inside the atmosphere \citep{pringle}. 
If this cooling takes place, direct accretion at a rate $\dot{M}_{\rm capt}$ 
onto the NS surface is permitted.  
 
\item{\it The direct accretion regime:} 
R$_{\rm M}$$<$R$_{\rm a}$, R$_{\rm co}$, $\dot{M}_{\rm w}$$>$$\dot{M}_{\rm lim}$. 
If R$_{\rm M}$$<$R$_{\rm co}$ and matter outside the 
magnetosphere cools efficiently, 
accretion onto the NS takes place at the same rate  
$\dot{M}_{\rm capt}$ (see Eq.~\ref{eq:dotmcapt}) at which it 
flows towards the magnetosphere. 
The corresponding luminosity is 
\begin{equation}
L_{\rm acc}= G M_{\rm NS} \dot{M}_{\rm capt} / R_{\rm NS} = 
2\times10^{35} \dot{M}_{-6} a_{\rm 10d}^{-2} 
v_8^{-4} ~{\rm erg ~s}^{-1}\simeq2\times10^{35}\dot{M}_{15}~{\rm erg ~s}^{-1}, 
\label{eq:lacc} 
\end{equation}
where $\dot{M}_{15}$=$\dot{M}_{\rm capt}$/10$^{15}$ g s$^{-1}$. 
This is the standard accretion regime; the system achieves the highest mass to 
luminosity conversion efficiency.  
\end{itemize} 

\section{Transitions and paths across different regimes}
\label{sec:trans}

We explore here the conditions under which 
transitions across different regimes take place. 
As emphasised in \S~\ref{sec:model}, these transitions occur 
when the relative positions of R$_{\rm M}$, R$_{\rm a}$, and R$_{\rm co}$
change in response to 
variations in the stellar wind parameters. In the following, 
since R$_{\rm M}$ depends only weakly on the orbital period and the 
total mass of the system, we fix a$_{\rm 10d}$=1, and 
investigate variations in the other four parameters: 
$\mu_{33}$, P$_{\rm s3}$, v$_{8}$, and $\dot{M}_{-6}$. 
The equations that define the conditions for transitions between different 
regimes are 
\begin{eqnarray}
R_{\rm M}>R_{\rm a} \Rightarrow \dot{M}_{-6}\lesssim0.45 \mu_{33}^2 v_{8}^{11} a_{\rm 10d}^2 & & {\it magnetic~barrier;} \\  
R_{\rm M}>R_{\rm co} \Rightarrow P_{\rm s3}\lesssim2.6\dot{M}_{-6}^{-1/4} 
v_{8}^{-1/4} a_{\rm 10d}^{1/2} \mu_{33}^{1/2} & & {\it centrifugal~barrier~with~R_{\rm M}>R_{\rm a};} \\ 
R_{\rm M}>R_{\rm co} \Rightarrow P_{\rm s3}\lesssim1.8 a_{\rm 10d}^{2/3}
\dot{M}_{-6}^{-1/3} v_{8}^{2/3} \mu_{33}^{2/3} & & {\it centrifugal~barrier~with~R_{\rm M}<R_{\rm a};} \\ 
P_{\rm s3}\gtrsim4.5 \dot{M}_{-6}^{-15/21} a_{\rm 10d}^{30/21} v_{8}^{60/21} \mu_{33}^{16/21}& &  {\it subsonic~propeller\rightarrow direct~accretion.}    
\label{eq:conditions}
\end{eqnarray} 
\begin{figure}[t!]
\centering
\includegraphics[height=5 cm]{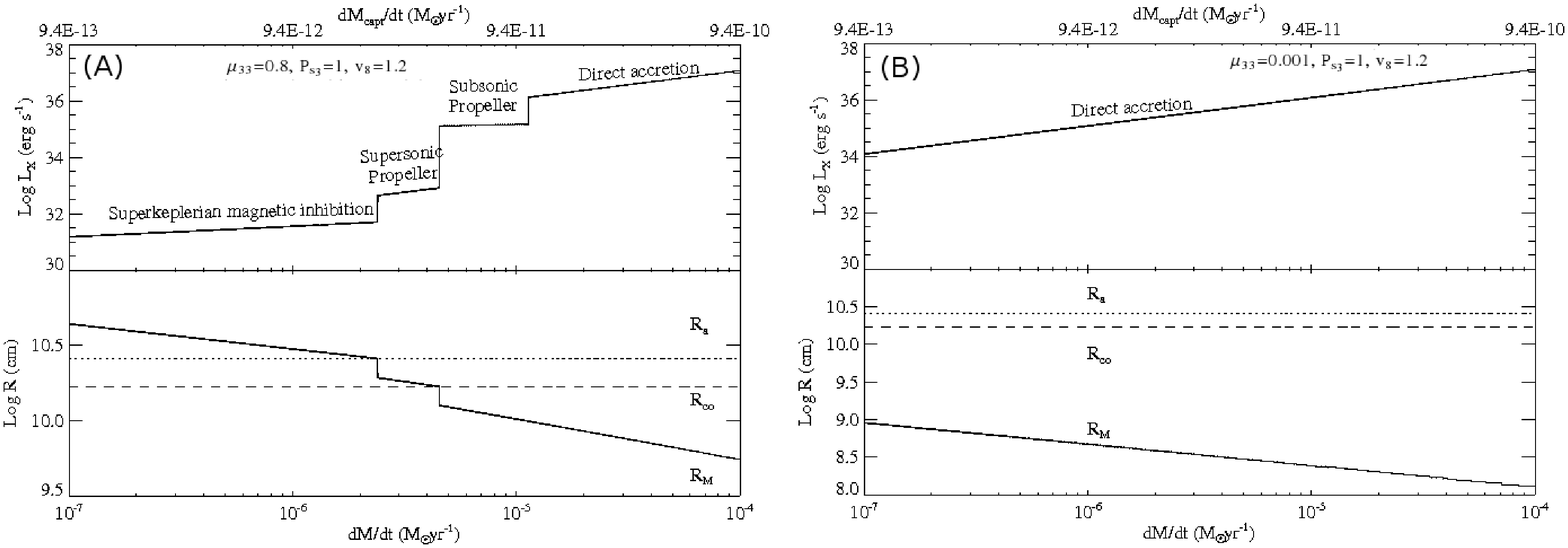}
\smallskip
\caption{\scriptsize (A) \textit{Upper panel}: Variation of the luminosity 
through different regimes, as a function of the mass loss rate from 
the companion star. 
In this case the parameters of the model are fixed at $\mu_{33}$=0.8, 
P$_{\rm s3}$=1, and v$_{8}$=1.2. 
\textit{Lower panel}: Relative position of the magnetospheric radius, 
R$_{\rm M}$ (solid line), with respect to the accretion radius R$_{\rm a}$ 
(dotted line), and the corotation radius R$_{\rm co}$ (dashed line), 
as a function of the mass loss rate from the companion star. \newline
(B) \textit{Upper panel}: Variation of the luminosity 
through different regimes, as a function of the mass loss rate 
from the companion star. 
In this case the parameters of the model are fixed at 
$\mu_{33}$=0.001, P$_{\rm s3}$=1, and v$_{8}$=1.2.  
\textit{Lower panel}: Relative position 
of the magnetospheric radius, R$_{\rm M}$ (solid line), 
with respect to the accretion radius R$_{\rm a}$ (dotted line), 
and corotation radius R$_{\rm co}$ (dashed line), 
as a function of the mass loss rate from the companion star.}  
\label{fig:magn_e_non} 
\end{figure}   
As an example, in Fig.~\ref{fig:magn_e_non}A we use the above equations 
to compute the luminosity swings for a system with  
$\mu_{33}$=0.8, P$_{\rm s3}$=1, and v$_{8}$=1.2. 
The lower panel of this figure shows that, for 0.1$<$$\dot{M}_{-6}$$<$100, 
the magnetospheric radius crosses both the centrifugal (R$_{\rm co}$) and magnetic 
(R$_{\rm a}$) barriers. Correspondingly, the system moves from  
the superKeplerian magnetic inhibition regime, to the supersonic and 
subsonic propeller regime, and, finally, to the direct accretion regime, 
giving rise to a six-decade luminosity swing from $\sim$10$^{31}$ to 
$\sim$10$^{37}$~erg~s$^{-1}$. 
We note that a large part of this swing (about five decades) is attained  
across the transitions from the superKeplerian magnetic inhibition to
the direct accretion regimes, 
which take a mere factor of $\sim$5 variation of $\dot{M}_{\rm w}$. 

\begin{figure}[t!]
\centering
\includegraphics[height=5 cm]{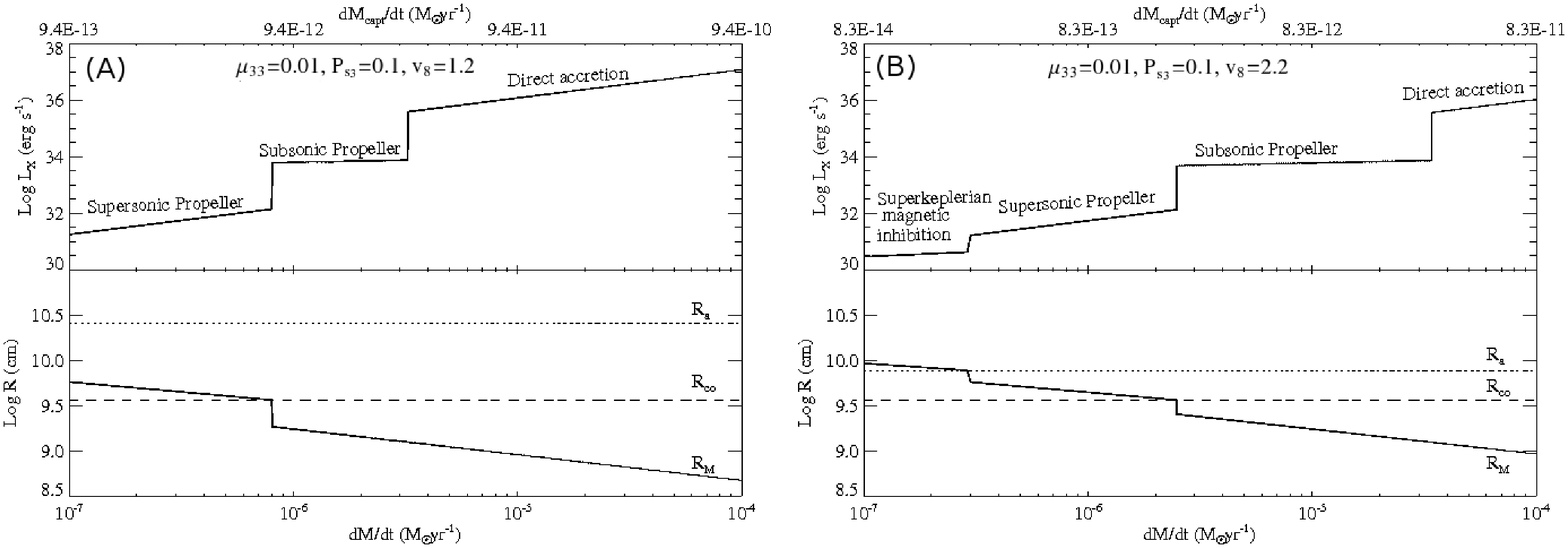}
\smallskip
\caption{\scriptsize (A) \textit{Upper panel}: Variation of the luminosity through 
different regimes, as a function of the mass loss rate. 
In this case the parameters of the model are fixed at 
$\mu_{33}$=0.01, P$_{\rm s3}$=0.1, and v$_{8}$=1.2.  
\textit{Lower panel}: Relative position of the 
magnetospheric radius, R$_{\rm M}$ (solid line), with respect to the 
accretion radius R$_{\rm a}$ (dotted line), and corotation radius 
R$_{\rm co}$ (dashed line), as a function of the mass loss rate from 
the companion star. 
(B) Same as (A) but for $\mu_{33}$=0.01, P$_{\rm s3}$=0.1, and 
v$_{8}$=2.2.} 
\label{fig:lshortspin} 
\end{figure}

Fig.~\ref{fig:magn_e_non}B  shows that, in the presence of a standard NS magnetic field ($10^{12}$~G), such abrupt luminosity jumps 
are not expected for a very slowly rotating (1000~s) NS 
(the other system parameters are the same 
as those of Fig.~\ref{fig:magn_e_non}A), since the magnetospheric 
radius is smaller than both R$_{\rm a}$ and R$_{\rm co}$, for any 
reasonable value of $\dot{M}_{\rm w}$. Therefore, 
the direct accretion regime applies, with  
the luminosity proportional to $\dot{M}_{\rm w}$. 

In Fig.~\ref{fig:lshortspin} we show the 
transitions for a system with $\mu_{33}$=0.01 and P$_{\rm s3}$=0.1.  
The wind velocity is v$_{8}$=1.2 in Fig.~\ref{fig:lshortspin}A, and v$_{8}$=2.2 
in Fig.~\ref{fig:lshortspin}B.  
These two figures show that, for sub-magnetar fields, a 100~s
spinning NS can undergo a transition across the magnetic barrier 
(besides the centrifugal barrier), for suitable parameters 
(a high wind velocity in the case at hand). Such transitions
take place over a more extended interval of mass loss rates. 
For instance Fig.~\ref{fig:lshortspin}B shows that an increase by  
a factor $\sim$100 in the mass loss rate is required, in this case,  
to achieve a factor $\sim$10$^5$ luminosity swing; this is  
comparable with the magnetar case of Fig.~\ref{fig:magn_e_non}A.\\ 
Taking into account of the examples discussed above, we conclude that:\\
- {\it Long spin period systems (P$_{\rm s3}$$\gtrsim$1) require magnetar-like 
B-fields ($\mu_{33}$$\gtrsim$0.1) in order for a large luminosity swing   
($\sim$10$^5$) to arise from modest variations in the wind parameters 
(e.g. a factor $\sim$5 in $\dot{M}_{-6}$). These luminosity swings might  
result from transition across different regimes through both the centrifugal 
and magnetic barriers.}\\ 
- {\it Shorter spin period systems (P$_{\rm s3}$$\ll$1) must posses lower magnetic 
fields ($\mu_{33}$$\ll$0.1) for similar transitions to take place.  
Somewhat larger variations in the wind parameters are required 
in order to achieve similar luminosity swings to those of the long period case, and 
transitions between different regimes occur in most cases through the centrifugal barrier.} \\
- {\it Few or no transitions are expected for systems with either high magnetic 
fields and short spin periods, or systems with lower magnetic fields and long spin periods. 
In the first case the centrifugal barrier halts the 
inflowing matter at R$_{\rm M}$ and accretion does not take place; such systems might 
thus be observable only at very low (X-ray) luminosity levels 
($\simeq$10$^{32}$-10$^{33}$~erg~s$^{-1}$). 
In the second case R$_{\rm M}$$<$R$_{\rm co}$ for a wide range of wind parameters,  
accretion can take place, and a high persistent luminosity is released  
($\simeq$10$^{35}$-10$^{37}$~erg~s$^{-1}$).} 

\section{Application to SFXT sources}
\label{sec:results}
 
\begin{figure*}[t!]
\centering 
\includegraphics[width=14.0 cm]{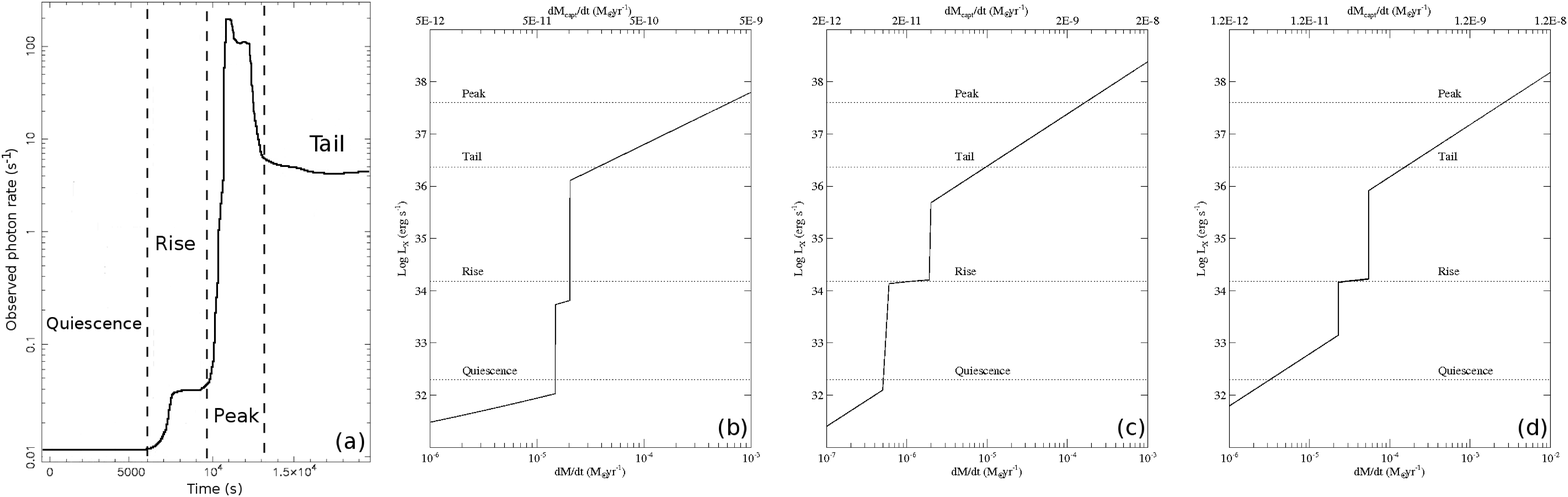}
\smallskip
\caption{ \scriptsize Application of the gated model to 
the \IGR\ transition from quiescence to outburst. 
Panel (a): A schematic representation of the \chan\ light curve 
of \IGR\ obtained by using the segments that were
not affected by pile-up \citep[see both panels of figure 2 in][]{zand05}. 
The different luminosity stages are clearly visible. 
According to \citet{zand05}, the count rates on the y-axis correspond 
to $2\times10^{32}$~erg~s$^{-1}$, 
$1.5\times10^{34}$~erg~s$^{-1}$, $2\times10^{36}$~erg~s$^{-1}$, and 
$4\times10^{37}$~erg~s$^{-1}$, in the quiescence state, the rise state, 
the tail, and the peak of the outburst, respectively. 
Panel (b): Interpretation of the quiescence to outburst transition of 
\IGR\ in terms of the magnetic barrier model.
The dotted horizontal lines mark the luminosity that 
divide the different regimes. The parameters of the model are fixed at 
$P_{\rm s3}$=1.3, $v_{8}$=1.4, and $\mu_{33}$=1. 
Panel (c): Interpretation of the quiescence to outburst transition of 
\IGR\ based on the centrifugal barrier model. The parameters of the model are fixed at 
$\mu_{33}$=0.08, $P_{\rm s3}$=0.4, and $v_{8}$=1. 
Panel (d): Same as panel (c) but here the parameters of the model are fixed at 
$\mu_{33}$=0.001, $P_{\rm s3}$=0.01, and $v_{8}$=2.} 
\label{fig:IGRJ17544} 
\end{figure*}

In this section we propose that transitions across different regimes 
caused by relatively mild variations of the wind parameters are responsible
for the outbursts of SFXTs. 
As a case study we consider \IGR\ \citep{sunyaev03}, a SFXT observed by \chan\ 
during a complex transition to and from a $\sim$1~hour-long outburst,
yielding the first detailed characterization of a SFXT light curve over a wide range of luminosity. 
The spin period of \IGR\ is presently unknown. 
\citet{zand05} showed that four different 
stages, with very different luminosity levels, 
could be singled out during the \chan\
observation: (a) a quiescent state with 
L$_{\rm X}$$\simeq$2$\times$10$^{32}$~erg~s$^{-1}$, (b) a rise stage with 
L$_{\rm X}$$\simeq$1.5$\times$10$^{34}$~erg~s$^{-1}$, (c) the outburst peak 
with L$_{\rm X}$$\simeq$4$\times$10$^{37}$~erg~s$^{-1}$, and (d) a post-outburst 
stage (or``tail'') with L$_{\rm X}$$\simeq$2$\times$10$^{36}$~erg~s$^{-1}$ 
\citep[see panel (a) of Fig.~\ref{fig:IGRJ17544}; these luminosities are for a source 
distance of $\sim$3.6 kpc,][]{rahoui08}.   
The maximum luminosity swing observed across these stages was 
a factor of $\gtrsim$6.5$\times$10$^4$. 

Motivated by the evidence for  $>$1000~s periodicities  
in \xte\ and \igg,\ we discuss first 
the possibility that \IGR\ contains a very slowly spinning NS. 
We use $\mu_{33}$=1, P$_{\rm s3}$=1.3, v$_{8}$=1.4, 
and show in Fig.~\ref{fig:IGRJ17544}(b) the different regimes 
experienced by such a NS as a function of the mass loss rate. 
For $\dot{M}_{-6}$$<$20 the above values give R$_{\rm M}$$>$R$_{\rm a}$ 
and R$_{\rm M}$$>$R$_{\rm co}$, such that superKeplerian magnetic inhibition of accretion applies.  
The expected luminosity 
in this regime, $\sim$10$^{31}$~erg~s$^{-1}$, is likely outshined by the X-ray luminosity 
of the supergiant star \citep[the companion star's luminosity is not shown in 
Fig.~\ref{fig:IGRJ17544}, but it is typically of order $\sim$10$^{32}$~erg~s$^{-1}$,][]
{cassinelli81}. We conclude that the lowest emission state 
(quiescence) of \IGR\ can be explained in this way, with the companion 
star dominating the high energy luminosity \citep{zand05}. 
The rise stage is in good agreement with the subKeplerian magnetic inhibition regime, 
where the luminosity ($\sim$10$^{34}$~erg~s$^{-1}$) is dominated by accretion of matter 
onto the NS due to the KHI. The uncertainty in the value of $h$ translates into 
an upper limit on the luminosity in this regime which is
a factor of $\sim$10 higher than that given above \citep{bozzo08}. 
During the outburst peak the direct accretion regime must apply at a  
mass loss rate of $\dot{M}_{-6}$$=$500. In this interpretation 
direct accretion must also be at work in the outburst tail at 
$\dot{M}_{-6}$$\sim$3, where  
a slight decrease in $\dot{M}_{\rm w}$ would cause the magnetic barrier to close and the 
source to return to quiescence. According to this interpretation, 
if \IGR\ has a spin period of $>$1000~s, then it must host a magnetar. 

Panel (c) of Fig.~\ref{fig:IGRJ17544} shows an alternative interpretation of 
the \IGR\ light curve, where we fixed $\mu_{33}$=0.08, P$_{\rm s3}$=0.4, and 
v$_{8}$=1. 
For this somewhat faster spin (and lower magnetic field), the luminosity variation is mainly 
driven by a transition across the centrifugal barrier (as opposed to the magnetic barrier). 
In this case, the quiescent state corresponds to the supersonic propeller regime 
($\dot{M}_{-6}$ $<$0.6), the rise stage to the subsonic propeller (0.6$<$$\dot{M}_{-6}$$<$2),  
while both the peak of the outburst and tail take place in the direct accretion regime 
at $\dot{M}_{-6}$=200 and $\dot{M}_{-6}$=10, respectively. 
Assuming an even faster NS spin period for \IGR,\ a weaker magnetic field would be
required.   
In panel (d) of Fig.~\ref{fig:IGRJ17544}, we show the results obtained by adopting 
$\mu_{33}$=0.001, P$_{\rm s3}$=0.01, and v$_{8}$=2. The $\sim$10$^{34}$~erg~s$^{-1}$ 
luminosity in the subsonic propeller regime compares well with the luminosity in the 
rise stage. However, the luminosity of the supersonic propeller regime 
is now significantly higher than the quiescence luminosity 
of $\sim$10$^{32}$~erg~s$^{-1}$ (this is a consequence of the higher value of 
$\dot{M}_{\rm w}$ for which the supersonic propeller regime is attained in this interpretation). 
We note that, the whole luminosity swing takes place 
for a wider range of mass loss rates, and the outburst peak luminosity requires 
$\dot{M}_{-6}$$\simeq$3000, an extremely high value even for an OB supergiant. 

Interpreting the properties of \IGR\ in terms of a NS 
with a spin periods $\ll$100~s is more difficult.   
For instance, for the subsonic propeller regime to set in, 
the mass loss rate corresponding to the transition across
R$_{\rm M}$=R$_{\rm co}$ must be lower than the limit 
fixed by Eq.~\ref{eq:dotmlimsub}. If instead the transition  
takes place at higher mass loss rate, the system goes directly from the supersonic 
propeller to the direct accretion regime (or vice versa), bypassing the subsonic propeller: 
therefore, the rise stage would remain unexplained. 
Since fast rotating NSs require lower magnetic fields for direct accretion to take place
while in outburst, Eq.~\ref{eq:dotmlimsub} is satisfied 
only for very high wind velocities (v$_{8}$$>$2-3).  
On the other hand, an increase by a factor of $\sim$2 in the wind velocity 
(with respect to the longer spin period solutions) would give a substantially lower 
$\dot{M}_{\rm capt}$, such that the subsonic and the direct accretion regime 
luminosities fall shortwards of the observed values (unless unrealistically high 
mass loss rate are considered).  
{\it Based on the above discussion, we conclude that \IGR\ 
likely hosts a slowly rotating NS, with spin period $>$100~s. 
Whether the magnetic barrier or the centrifugal barrier sets in, causing 
inhibition of accretion away from the outbursts, will depend on whether 
the spin period is longer or shorter then $\sim$1000~s.}   

As another example we discuss the case of \igrjj,\ a SFXTs with  
a spin period of 228~s. The luminosity behaviour  
of this source is still poorly known. An outburst
at 5$\times$10$^{36}$~erg~s$^{-1}$ was observed with \int\  
\citep[assuming a distance of 12.5 kpc,][]{Lutovinov05,smith04}, which did not 
detect the source before the outburst down to a level 
of 5$\times$10$^{35}$~erg~s$^{-1}$. About a week later,  
\xmm\ revealed the source at 5$\times$10$^{34}$~erg~s$^{-1}$  
and discovered the 228~s pulsations \citep{Lutovinov05}. 
If the direct accretion regime applied all the way to the 
lowest luminosity level observed so far, then an upper limit 
of $\mu_{33}$$\simeq$0.004 would be obtained by imposing that 
the NS did not enter  
the subsonic propeller regime. 
On the other hand, if the luminosity measured by 
\xmm\ signalled that the source entered the subsonic propeller
regime, while direct accretion occurred only during the outburst 
detected by \int,\ then a considerably higher 
magnetic field of $\mu_{33}$$\simeq$0.07 would be required. 

{\it The above discussion emphasizes the importance of determining, 
through extended high sensitivity observations,  
the luminosity at which transitions between different 
source states occur, in particular the lowest luminosity 
level for which direct accretion is still at work. 
In combination with the NS spin, this can
be used to infer the NS magnetic field.  
Alternatively accretion might take place unimpeded 
at all luminosity levels of SFXTs, a possibility
which requires a very clumpy wind as envisaged in 
other scenarios \citep{walter07}. In this case 
the NS magnetic field can be considerably lower 
than discussed here.} 

\section{Conclusions}
\label{sec:conclusions}

In this paper we reviewed the theory of wind accretion in HMXBs hosting a magnetic
NS with a supergiant companion, and considered in some detail the 
interaction processes between the inflowing plasma and the magnetosphere, 
that are expected to take place when direct accretion onto the 
NS surface is inhibited. We then applied this theory to SFXTs   
and showed that their large luminosity swings
between quiescence and outburst (up to a factor of $\sim$10$^{5}$)
can be attained in response to relatively modest variations of the 
wind parameters, provided the system undergoes transitions across different 
regimes. We found that such transitions can be driven mainly by 
either: (a) a centrifugal barrier mechanism, which halts direct accretion 
when the NS rotation becomes superKeplerian at the magnetospheric 
radius, or (b) a magnetic barrier mechanism, when the magnetosphere 
extends beyond the accretion radius. Which mechanism  
and wind interaction regime apply will depend  
sensitively on the NS spin period and magnetic field, 
besides the velocity and mass loss rate in the supergiant's wind. 
In particular, the magnetic barrier mechanism requires 
long spin periods ($\gtrsim$1000~s) coupled with magnetar-like 
fields ($\gtrsim$10$^{14}$~G). 
On the other hand, magnetar-like 
fields would also be required if the centrifugal barrier sets in  
at relatively high luminosities ($\gtrsim 10^{36}$~erg~s$^{-1}$)  
in NSs with spin periods of hundreds seconds. 

Evidence has been found that the spin periods of a few SFXTs 
might be as long as 1000-2000~s. Motivated by this, we 
presented an interpretation of the 
activity of \IGR\ (whose spin period is unknown) in terms of the   
magnetic barrier by a 1300~s spinning NS  
and showed that the luminosity
stages singled out in a \chan\ observation of this source 
are well matched by the different regimes of wind-magnetosphere
interaction expected in this case.
We discussed also an interpretation  
of this source based on the centrifugal barrier and a slightly  shorter 
spin period (400~s), which can reproduce the luminosity stages comparably well. 
We emphasise that in both solutions the required magnetic field 
strength ($\gtrsim$10$^{15}$~G and $\gtrsim$8$\times$10$^{13}$~G, respectively) 
are in the magnetar range. 

{\it While the possibility that magnetars are hosted in binary system with supergiant 
companions has been investigated by several authors \citep[e.g.,][]{zhang04,liu},   
clear observational evidence for such extremely high magnetic field NS 
in binary systems is still missing. According to the present study, 
long spin period SFXTs might provide a new prospective for detecting and studying 
magnetars in binary systems.}  

\acknowledgments
EB thanks DTU Space, Technical University of Denmark and IASF-INAF for 
grant support for this conference. 
This work was supported through ASI and MIUR grant.


\begin{thebibliography}{99}

\bibitem[\protect\citeauthoryear{Bozzo et al.}{2008}]{bozzo08} 
Bozzo, E., Falanga, M., Stella, L. 2008, ApJ, 683, 1031 

\bibitem[\protect\citeauthoryear{Burnard et al.}{1983}]{bur} 
Burnard, D. J., Arons, J., \& Lea, S. M. 1983, ApJ, 266, 175 
	
\bibitem[\protect\citeauthoryear{Cassinelli et al.}{1981}]{cassinelli81} 
Cassinelli, J. P., et al. 1981, ApJ, 250, 677 

\bibitem[\protect\citeauthoryear{Davidson \& Ostriker}{1979}]{davidson}
Davidson, K. \& Ostriker, J. P. 1979, ApJ, 179, 585 

\bibitem[\protect\citeauthoryear{Davies et al.}{1979}]{davies} 
Davies, R. E., Fabian, A. C., \& Pringle, J. E. 1979, MNRAS, 186, 779  

\bibitem[\protect\citeauthoryear{Davies \& Pringle}{1981}]{pringle} 
Davies, R. E. \& Pringle, J. E. 1981, MNRAS, 196, 209 

\bibitem[\protect\citeauthoryear{Elsner \& Lamb}{1977}]{elsner1977} 
Elsner, R. F. \& Lamb, F. K. 1977, ApJ, 215, 897 

\bibitem[\protect\citeauthoryear{Frank et al.}{2002}]{fkr} 
Frank J., et al. 2002, Accretion Power in Astrophysics 
(3th ed., Cambridge University Press) 

\bibitem[\protect\citeauthoryear{Harding et al.}{1992}]{hard} 
Harding, A. K. \& Leventhal, M. 1992, Nature, 357, 388 

\bibitem[\protect\citeauthoryear{Ikhsanov \& Pustil'nik}{1996}]{ikhsanov} 
Ikhsanov, N. R. \& Pustil'nik, L. A. 1996, A\&A, 312, 338 

\bibitem[\protect\citeauthoryear{Illarionov \& Sunyaev}{1975}]{illarionov75} 
Illarionov, A. F. \& Sunyaev, R. A. 1975, A\&A, 39, 185  

\bibitem[\protect\citeauthoryear{In't Zand}{2005}]{zand05}
In 't Zand 2005, A\&A, 441, L1 

\bibitem[\protect\citeauthoryear{Lipunov}{1992}]{lipunov02}
Lipunov, V. M. 1992, Astrophysics of Neutron Stars (Springer-Verlag) 

\bibitem[\protect\citeauthoryear{Liu \& Yan}{2006}]{liu}
Liu, Q. Z. \& Yan, J. Z. 2006, AdSpR, 38, 2906 

\bibitem[\protect\citeauthoryear{Lutovinov et al.}{2005}]{Lutovinov05}
Lutovinov, A., et al. 2005, A\&A, 444, 821 

\bibitem[\protect\citeauthoryear{Mori \& Ruderman}{2003}]{mori} 
Mori, K. \& Ruderman, M. A. 2003, ApJ, 592, L75 

\bibitem[\protect\citeauthoryear{Negueruela et al.}{2008}]{negueruela08}
Negueruela, I., et al. 2008, in press [astro-ph/0801.3863] 

\bibitem[\protect\citeauthoryear{Rahoui et al.}{2008}]{rahoui08} 
Rahoui, F., Chaty, S., Lagage, P.-O., Pantin, E. 2008, A\&A, 484, 801  

\bibitem[\protect\citeauthoryear{Sguera et al.}{2005}]{sguera05} 
Sguera, V., et al. 2005, A\&A, 444, 221 

\bibitem[\protect\citeauthoryear{Sguera et al.}{2006}]{sguera06} 
Sguera, V., et al. 2006, ApJ, 646, 452 

\bibitem[\protect\citeauthoryear{Sidoli et al.}{2007}]{sidoli07} 
Sidoli, L., et al. 2007, A\&A, 476, 1307  

\bibitem[\protect\citeauthoryear{Smith}{2004}]{smith04} 
Smith, D. M. 2004, Astr. Tel., 338 

\bibitem[\protect\citeauthoryear{Stella et al.}{1986}]{stella86} 
Stella, L., White, N. E., \& Rosner, R. 1986, ApJ, 308, 669 

\bibitem[\protect\citeauthoryear{Sunyaev et al.}{2003}]{sunyaev03}  
Sunyaev, R. A., et al. 2003, Astr. Tel., 190 

\bibitem[\protect\citeauthoryear{Toropina et al.}{2006}]{toropina} 
Toropina, O. D., Romanova, M. M., \& Lovelace, R. V. E. 2006, MNRAS, 371, 569 

\bibitem[\protect\citeauthoryear{Walter et al.}{2006}]{walter06}
Walter, R., et al. 2006, A\&A, 453, 133

\bibitem[\protect\citeauthoryear{Walter \& Zurita Heras}{2007}]{walter07}
Walter, R. \& Zurita Heras, J. A. 2007, A\&A, 476, 335 

\bibitem[\protect\citeauthoryear{Zhang et al.}{2004}]{zhang04} 
Zhang, F., Li, X.-D. \& Wang, Z.-R. 2004, ChjAA, 4, 320

\end{thebibliography}
\end{document}